\documentstyle[aps,preprint,floats,epsfig]{revtex}

\begin{document}

\preprint{\tighten\vbox{\hbox{\hfil CLEO CONF 99-10}}}

\title{$\mathbf{b \rightarrow s \gamma}$ Branching Fraction and CP Asymmetry}

\author{CLEO Collaboration}
\date{\today}

\maketitle
\tighten

\begin{abstract} 
We present preliminary results on $b \rightarrow s \gamma$ from the
CLEO experiment. An updated result on the branching fraction is
reported at $\cal{B}$$(b \rightarrow s \gamma) = (3.15 \pm 0.35 \pm 0.32
\pm 0.26)\times 10^{-4}$, where the first uncertainty is statistical,
the second is systematic, and the third for model dependence. We also
describe a new analysis performed to search for CP asymmetry in $b
\rightarrow s \gamma$ decays. We observe no such asymmetry and set
conservative limits at $-0.09 <$ $\cal{A}$ $< 0.42$.
\end{abstract}
\newpage

\renewcommand{\thefootnote}{\fnsymbol{footnote}}

% Insert author and address list here
\begin{center}
S.~Ahmed,$^{1}$ M.~S.~Alam,$^{1}$ S.~B.~Athar,$^{1}$
L.~Jian,$^{1}$ L.~Ling,$^{1}$ A.~H.~Mahmood,$^{1,}$%
\footnote{Permanent address: University of Texas - Pan American, Edinburg TX 78539.}
M.~Saleem,$^{1}$ S.~Timm,$^{1}$ F.~Wappler,$^{1}$
A.~Anastassov,$^{2}$
% ...
\end{center}
 
\small
\begin{center}
$^{1}${State University of New York at Albany, Albany, New York 12222}\\
$^{2}${Ohio State University, Columbus, Ohio 43210}\\
% ...
\end{center}
 
\setcounter{footnote}{0}
\newpage

\section{Introduction}
Because flavor changing neutral currents are forbidden in the Standard
Model, electroweak penguins such as $b \rightarrow s \gamma$ give a
direct look at loop and box processes. The inclusive branching
fraction for $b \rightarrow s \gamma$ is important for restricting
physics beyond the Standard Model, and CLEO's 1995 published
result\cite{pubres} of $(2.32 \pm 0.57 \pm 0.35)\times 10^{-4}$
generated much theoretical interest\cite{theorint}. The branching
fraction result presented here is an improved measurement with 60\%
additional data and enhanced analysis techniques. The theoretical
branching fraction prediction from the Standard Model has also
improved, with a full next-to-leading-log calculation\cite{nllcalc} of
$(3.28 \pm 0.33)\times 10^{-4}$. We also introduce a new analysis: a
search for CP asymmetry in $b\rightarrow s \gamma$.

\section{The CLEO Detector}
Data for these analyses were taken with the CLEO detector at the
Cornell Electron Storage Ring (CESR) with center of mass energy at the
$\Upsilon\mbox{(4S)}$ resonance (10.58 GeV). Additional data were
taken 60 MeV below the $\Upsilon\mbox{(4S)}$ (off-resonance) for
continuum background subtraction. The CLEO detector\cite{cleodet}
measures charged particles over 95\% of 4$\pi$ steradians with a
system of cylindrical drift chambers. The barrel and endcap CsI
calorimeters cover 98\% of 4$\pi$, and the energy resolution for photons
near 2.5 GeV in the central angular region $(|\cos \theta_{\gamma} | <
0.7)$ is 2\%. In 1995, a silicon vertex detector replaced the inner
most tracking chamber, though it is not used in the analyses described
here.

\section{Branching Fraction Analysis}
This analysis is discussed in detail elsewhere\cite{bfa} and is
briefly covered here. The signal for $b \rightarrow s \gamma$ is a
photon from $B$ meson decay with $2.1 < E_\gamma < 2.7$~GeV (the
published analysis\cite{pubres} used $2.2 < E_\gamma < 2.7$~GeV).  The
Fermi momentum of the $b$ quark in the $B$ meson and the momentum of
the $B$ meson in the lab frame ($B$ mesons are produced from
$\Upsilon\mbox{(4S)}$ decays with approximately 300~MeV/$c$ of
momentum at CESR) Doppler broadens the photon line. Calculations
using a spectator model\cite{spectmod} indicate that 85-94\% of the
signal lies in the photon energy range.

To obtain a photon energy spectrum, we select hadronic events with a
high energy calorimeter cluster in the central region $(| \cos
\theta_\gamma | < 0.7)$. We reject a cluster if, when paired with
another calorimeter cluster in the event, it forms a combined $\gamma
\gamma$ mass consistent with a $\pi^0$ or $\eta$. We also require the
cluster shape be consistent with that of a single high energy photon.

A na\"{i}ve approach to this analysis would be to measure the photon
spectrum in on-resonance data and use the off-resonance data to
subtract the continuum background. But the backgrounds from continuum
with initial state radiation ($e^+ e^- \rightarrow q \bar{q} \gamma$)
(ISR) and from continuum processes ($e^+ e^- \rightarrow q \bar{q}$)
involving a high energy $\pi^0$, $\eta$, or $\omega$ where one of the
daughter photons is not detected dominate to such an extent that
extraction of a signal is impossible. We therefore suppress the
continuum background with two separate methods and subtract what
remains with off-resonance data.

The first continuum suppression scheme\cite{jae} involves exploiting
differences in the event shapes between signal ($B\bar{B}$) and
continuum. While continuum events appear jetty and continuum with ISR
appear jetty when transformed to the rest frame of the system after
the initial state radiation, the signal events are spherical due to
the other $B$ meson decaying in the event. We characterize the event
shape with eight variables[6]: the Fox-Wolfram second moment ($R_2$),
ratio of the sum of transverse momenta of particles more than
45$^\circ$ away from the high energy photon axis to the total scalar
sum of particle momenta in the event excluding the momentum of the
high energy photon ($S_\perp$), $R_2$ transformed to the rest frame of
the system after ISR (assuming the high energy photon in the event is
initial state radiation) ($R_2'$), $\cos \theta$ between the photon and
the thrust axis of the rest of the event transformed into the rest
frame of the system after ISR ($\cos \theta'$), and the energies of
particles found in 20$^\circ$ and 30$^\circ$ cones parallel and
anti-parallel to the high energy photon direction. Since no one
variable provides enough discriminating power, we combine them with a
neural network into a single variable $r$. The distribution of $r$
tends towards $-1$ for continuum and ISR events and $+1$ for signal $b
\rightarrow s \gamma$ events. This shape method is approximately 30\%
efficient for signal.

The other method of suppressing continuum and ISR events is a
pseudo-reconstruction of the parent $B$ meson involved in the $b
\rightarrow s \gamma$ decay. We look for a charged or neutral kaon
with up to four pions, one of which may be a $\pi^0$, which when
combined with the high energy photon are consistent with being the
decay products of a $B$ meson. We try all combinations of kaons and
pions given the above restrictions and choose the one that minimizes a
$\chi^2$ that includes,
\begin{equation}
\chi^2_B = \left(\frac{M-M_B}{\sigma_M}\right)^2 + 
           \left(\frac{E-E_{\text{beam}}}{\sigma_E}\right)^2
\end{equation}
as well as contributions from particle identification such as $dE/dx$
and $\pi^0$ and $K^0_s$ mass deviations. In Eq. (1), $M =
\sqrt{E^2_{\text{beam}} - P^2}$ is the beam-constrained mass, where
$P$ is the vector sum of particle momenta comprising the $B$ meson
candidate, $M_B$ is the $B$ mass (5.28 GeV/c$^2$), $E$ is the energy of
the candidate $B$, and $\sigma_M$ and $\sigma_E$ are the resolutions
on the beam constrained mass and the measured $B$ energy
respectively. Events that have a minimum $\chi^2_B < 20$ are considered
pseudo-reconstructed (``pseudo'' is used because this method is only for
continuum suppression, and we are not concerned that the parent $B$
reconstruction be exactly correct). If the minimum $\chi^2_B > 20$ or not one
combination of kaons and pions was found in the event, then the shape
analysis described above is used.  

For events that are pseudo-reconstructed, we calculate the angle
between the thrust axis of the inferred signal B meson and the thrust
axis of the rest of the event $(\cos \theta_{tt})$. Events that are in
actuality continuum typically have $|\cos \theta_{tt}|$ peaked at 1,
while signal events have a flat distribution. To get the most
discriminating power, we combine $\chi^2_B$ and $\cos \theta_{tt}$
from the pseudo-reconstruction and $r$ from the shape neural network
into a new variable $r_c$ using another neural network. As with $r$, the
distribution of $r_c$ tends towards $+1$ for signal events and $-1$ for
continuum. While this combined method is 10\% efficient for signal, it
reduces the continuum background by an additional factor of four over
the shape method alone.  

We then weight each event according to its value of $r_c$ if it is
pseudo-reconstructed or $r$ if not. The weighting scheme\cite{bfa} is
designed so that the weighted yield is equivalent to the event yield
in the absence of continuum backgrounds. This procedure gives the
smallest statistical uncertainty on this background subtracted
yield. Note that there are non-continuum backgrounds discussed below,
but these are small compared to continuum.  The continuum background
remaining after the suppression efforts is determined from
off-resonance data.

$B\bar{B}$ backgrounds are dominated by $b \rightarrow c$ ({\em e.g.}
$B \rightarrow X \pi^0$ and $B \rightarrow X \eta$) though $b
\rightarrow u$ and $b \rightarrow s g$ may have a small contribution.
We estimate the $B \bar{B}$ backgrounds with Monte Carlo (MC), but
correct the $\pi^0$ and $\eta$ momentum spectra for any differences
between data and MC. We obtain the momenta spectra from data by
treating $\pi^0$'s and $\eta$'s as if they were photons and follow the
same analysis procedure using weights described above. Thus, the MC is
used only for the $\pi^0$ and $\eta$ veto efficiencies and for any
other small $B\bar{B}$ backgrounds not originating from missing a
$\gamma$ daughter from a $\pi^0$ or $\eta$.

\begin{figure}
\epsfig{file=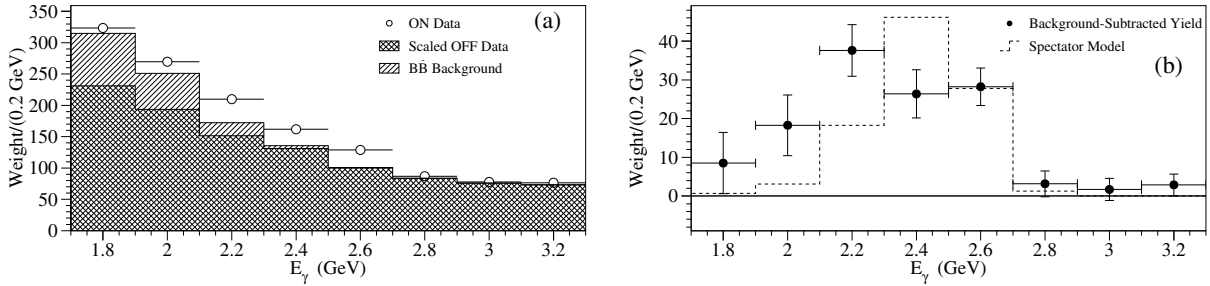, width=6.5in}
\caption{Photon energy spectra. (a) On-resonance data (points),
luminosity scaled off-resonance data (double hatched), and $B\bar{B}$
background (single hatched). (b) Background-subtracted data (points)
and Monte Carlo prediction of the shape of the $b \rightarrow s
\gamma$ signal from a spectator model calculation ($<m_b> = 4.88
\text{ GeV}/c^2$) and Fermi momentum set at 250 MeV/$c$. Only
statistical uncertainties are shown.}
\end{figure}

The weighted yields from 3.1~fb$^{-1}$ of on-resonance data plotted
with respect to photon energy are shown in Fig.\ 1, with all of the
backgrounds shown in (a) and the background subtracted weighted yield,
which is the signal, shown in (b). 1.6~fb$^{-1}$ of off-resonance data
were used for the continuum subtraction. The expected spectrum for $b
\rightarrow s \gamma$ from a spectator model is also shown in (b). In
the region of interest ($2.1 < E_\gamma < 2.7$~GeV) we measure $500.5
\pm 7.4$ on-resonance weighted events, $382.3 \pm 7.0$ scaled
off-resonance weighted events, and estimate the $B\bar{B}$ background
to be 20.6 weighted events (all uncertainties statistical). Applying
the $\pi^0$ and $\eta$ corrections to the estimate ($1.9 \pm 0.5$ and
$3.6 \pm 1.1$ weighted events respectively), we derive the weighted,
background-subtracted yield of $92.2 \pm 10.3 \pm 6.5$, where the
first uncertainty is statistical and the second is systematic.

For a typical sample of MC $b \rightarrow s \gamma$ events,
approximately 50\% pass all analysis requirements and of these, 40\%
are pseudo-reconstructed ($\chi^2_B <20$). But since we use weighted
events for the signal yield, we need a weighted efficiency; the sum of
weights passing all requirements. Because of our definition of
weights, the weighted efficiency tends to be smaller than the
efficiencies stated above. To model the signal, we use the spectator
model of Ali and Greub\cite{spectmod}, which includes gluon
bremsstrahlung and other higher order radiative effects. Details of MC
production and parameters used are given in ref~\cite{pubres}. The
weighed efficiency is $(4.43 \pm 0.29 \pm 0.22 \pm 0.03 \pm
0.31)\times 10^{-2}$, where the first uncertainty is due to spectator
model inputs, the second due to recoil system hadronization, the third
due to uncertainty in the production ratio of neutral and charged $B$
meson pairs, and the last to detector modeling.

The weighted efficiency is combined with the background subtracted
weighted yield to give the inclusive branching fraction of $b
\rightarrow s \gamma$ as $(3.15 \pm 0.35 \pm 0.32 \pm 0.26)\times
10^{-4}$, where the uncertainties are statistical, systematic and for
model dependence respectively. Our results are in agreement with the
Standard Model prediction. Conservatively allowing for the systematic
uncertainty, we find that the branching fraction must be between
$2.0\times 10^{-4}$ and $4.5\times 10^{-4}$ (each limit at 95\% CL).  

\section{CP Asymmetry} 
The SM predicts no CP asymmetry in $b \rightarrow s \gamma$ decays,
but some recent theoretical work\cite{cp} suggests that non-SM physics may
significantly contribute to a CP asymmetry. Furthermore, if new
physics has a weak phase difference near 90$^{\circ}$ with respect to the
SM and the strong phase is non-zero, the new physics would only
slightly alter the $b \rightarrow s \gamma$ inclusive branching ratio,
but could produce a large effect in the asymmetry. Since we can reuse
much of the machinery from the branching fraction measurement, a
search for CP asymmetry is an obvious extension to the $b \rightarrow
s \gamma$ analysis.

The pseudo-reconstruction procedure described above is used to tag the
$b$ quark flavor ($b$ or $\bar{b}$), because it determines the
particles decayed from the signal $B$ meson. But since the
pseudo-reconstruction was designed for continuum suppression, we must
determine how often it reconstructs the $b$ flavor correctly. If a
signal $B$ meson decays to a neutral kaon and overall neutral pions
({\em e.g.} $B^0 \rightarrow K^0_s \pi^+ \pi^-$), then we cannot
determine the $b$ flavor from pseudo-reconstruction. But for the
majority of cases where the $b$ flavor is determinable, it can be
deduced by the charge of the kaon, or the sum pion charge if the kaon
is neutral. From MC, we estimate the probability of reconstructing the
incorrect $b$ flavor when the $b$ flavor is determinable to be $8.3
\pm 1.6 \%$.  The quoted uncertainty includes a combination of MC
statistics and uncertainties derived from varying spectator model
inputs to the MC and varying the ratio of charged to neutral $B$ meson
decays. We also take into account the probability of tagging the $b$
flavor when it should be indeterminable and vice--versa, though these
effects are small and will not be discussed further here.

The asymmetry we measure is $(N_1 - N_2)/(N_1 + N_2)$, where $N_1$ is
the weighted yield tagged as $b$ quarks from pseudo-reconstruction and
determinable as such, and $N_2$ is similarly the weighted yield tagged
as $\bar{b}$. From the misreconstruction rates described above, we
apply a multiplicative correction factor of $1.22 \pm 0.04$ to the
asymmetry. In our misreconstruction determination, we assumed that the
rate of mistagging a $b$ as a $\bar{b}$ was the same as mistagging a
$\bar{b}$ as a $b$.  Initial studies lead us to apply a conservative
additive systematic error on the asymmetry of 5\% to account for
unequal mistagging rates, though we observe no measurable asymmetry in
the off-resonance data and other control samples.  

Using the weighted yield from pseudo-reconstructed events with photons
within $(2.2 < E_\gamma < 2.7\text{ GeV})$, we observe $36.8 \pm 5.2$ weights
reconstructed as $b$ quarks and $28.4 \pm 5.2$ weights reconstructed
as $\bar{b}$ quarks (uncertainties are statistical only), leading to a
raw asymmetry measurement of $0.13 \pm 0.11$. Applying the correction
factors, we obtain our preliminary CP asymmetry measurement in $b
\rightarrow s \gamma$ decays: $(0.16 \pm 0.14 \pm 0.05)\times(1.0 \pm 0.04)$,
where the first number is the central value, the second is the
statistical uncertainty, the third is the additive systematic
described above, and the multiplicative factor is from the uncertainty
on the mistagging rate correction. Within the uncertainties, we
measure no CP asymmetry. From this result, we derive 90\% CL limits on
the CP asymmetry ($\cal{A}$) of $-0.09 < $ $\cal{A}$ $ < 0.42$.

\section{Summary and Outlook}
We presented CLEO's preliminary results on the $b \rightarrow s
\gamma$ branching ratio and the CP asymmetry in $b \rightarrow s
\gamma$ decays with approximately 3.1~fb$^{-1}$ of on-resonance
data. CLEO has since accumulated a total of 10~fb$^{-1}$ of
on-resonance data. Work is underway to determine the $b \rightarrow s
\gamma$ inclusive branching fraction and CP asymmetry with all of the
data. We are also exploring extending both $b \rightarrow s \gamma$
analyses by tagging the $b$ flavor with a lepton from the other $B$
meson in the event it undergoes semileptonic decay.  This procedure
offers further continuum suppression and may improve the asymmetry
measurement by allowing the use of events where the $b$ flavor is
undeterminable by pseudo-reconstruction.

We gratefully acknowledge the effort of the CESR staff in providing us
with excellent luminosity and running conditions. We also thank M.\
Neubert for bringing the theoretical interest in $b \rightarrow s
\gamma$ CP asymmetry to our attention. J.R.\ Patterson and I.P.J.\
Shipsey thank the NYI program of the NSF, M.\ Selen thanks the PFF
program of the NSF, M.\ Selen and H.\ Yamamoto thank the OJI program
of the DOE, J.R.\ Patterson, K.\ Honscheid, M.\ Selen and V.\ Sharma
thank the A.P.\ Sloan Foundation, M.\ Selen and V.\ Sharma thank the
Research Corporation, F.\ Blanc thanks the Swiss National Science
Foundation, and H.\ Schwarthoff and E.\ von Toerne thank the Alexander
von Humboldt Stiftung for support. This work was supported by the
National Science Foundation, the U.S.\ Department of Energy, and the
Natural Sciences and Engineering Research Council of Canada.

\end{document}